\documentclass[12pt]{article}
\usepackage{fullpage}
\usepackage[utf8]{inputenc}
\usepackage{psfrag}
\usepackage{amssymb}
\usepackage{amsmath}
\usepackage[dvips]{graphicx}
\usepackage{url}
\usepackage{multirow}
\newcommand{\seq}[1]{\left\langle #1\right\rangle}

\newcommand{\set}[1]{\left\{ #1\right\}}
\newcommand{\gilt}{:}

\newcommand{\real}{\mathbb{R}}

\newcommand{\realrange}[2]{\left[#1, #2\right]}

\newcommand{\unitrange}[2]{\realrange{0}{1}}

\newcommand{\llabel}[1]{\label{\labelprefix:#1}}
\newcommand{\labelprefix}{} 

\newcommand{\discussionsize}{\small}

\newcommand{\notiz}[1]{}
\newcommand{\frage}[1]{}

\newenvironment{code}{\noindent
\begin{tabbing}%
\hspace{2em}\=\hspace{2em}\=\hspace{2em}\=\hspace{2em}\=\hspace{2em}\=%
\hspace{2em}\=\hspace{2em}\=\hspace{2em}\=\hspace{2em}\=\hspace{2em}\=%
\kill}{\end{tabbing}}

\newcommand{\labelcommand}{}
\newcommand{\captiontext}{}
\newsavebox{\codeparam}
\newcounter{lineNumber}
\newenvironment{disscodepos}[3]{%
\renewcommand{\labelcommand}{#2}%
\renewcommand{\captiontext}{#3}%
\sbox{\codeparam}{\parbox{\textwidth}{#3}}%
\begin{figure}[#1]\begin{center}\begin{code}\setcounter{lineNumber}{1}}{%
\end{code}\end{center}\caption{\llabel{\labelcommand}\captiontext}\end{figure}}

{\end{disscodepos}}

\newdimen\endofsize\endofsize=0.5em
\def\endofbeweis{~\quad\hglue\hsize minus\hsize
                 \hbox{\vrule height \endofsize width
\endofsize}\par}
\newenvironment{proof}{Proof:}{\endofbeweis}

\newcommand{\ignore}[1]{}

\newcommand{\Gup}{G_{\uparrow}}
\newcommand{\Gdown}{G_{\downarrow}}

\newcommand{\Eup}{E_{\uparrow}}
\newcommand{\Emarked}{E_{\mathrm{marked}}}

\newcommand{\concat}{\mathbf{*}} 

\newtheorem{theorem}{Theorem}

\title{Time Dependent Contraction Hierarchies\\ --- Basic Algorithmic Ideas\thanks{Partially supported by
DFG grant SA 933/4-1 and a Google Research Award}}
\author{Veit Batz, Robert Geisberger and Peter Sanders\\\normalsize
Universit\"at Karlsruhe (TH), 76128 Karlsruhe, Germany\\\normalsize {\tt
\{batz,robert.geisberger,sanders\}@ira.uka.de}}
\begin{document}

\maketitle

\begin{abstract}
Contraction hierarchies are a simple hierarchical routing technique
that has proved extremely efficient for static road networks.
We explain how to generalize them to networks with time-dependent edge weights.
This is the first hierarchical speedup technique for time-dependent routing that
allows bidirectional query algorithms.
\end{abstract}

\section{Introduction}

This technical note explains how contraction hierarchies (CHs) can be
generalized to allow time-dependent edge weights. 
We assume familiarity with CHs
\cite{GSS08b,Gei08}. Like many of the most successful speedup techniques for
routing in road networks, the CH query-algorithm uses
\emph{bidirectional} search.  This is a challenge since bidirectional
searching in a time-dependent network requires knowing the arrival
time\footnote{Wlog we assume that a query specifies source,
destination and departure time.} which is what we want to compute in
the first place.


Due to the difficulty of bidirectional routing, the first promising
approaches to fast routing used goal directed rather than hierarchical
routing and accepted suboptimal routes \cite{NDLS08}.  SHARC routing \cite{BD08}
was specifically developed to encode hierarchical information into a
goal-directed framework allowing unidirectional search and recently
was generalized to exact time-dependent routing \cite{Del08}.
Schultes \cite{Sch08} gives a way to make queries in static
networks unidirectional but this approach does not directly yield 
a time-dependent approach.

\section{Preliminaries}\label{s:preliminaries}

There are classical results on time-dependent route planning
\cite{CH66}  that show that a simple generalization of Dijkstra's
unidirectional algorithm works for time-dependent networks $G=(V,E)$ if the
objective function is travel time and a cost function
$f:\real\rightarrow\real$ has the \emph{FIFO-property}:
$\forall \tau<\tau'\gilt \tau+f(\tau)\leq \tau'+f(\tau')$, i.e., there is no
overtaking. We focus on this case and further assume that the travel
time functions are representable by a piece-wise linear
function. However, all our algorithms view travel-time functions (TTFs) as an
abstract data type with a small number of operations, basically
evaluation, chaining (operation $\concat$ computes a time-dependent function for a sequence of edges) and minimum computations. Also note, that the
format used in public transportation with lists of departure times and
arrival times can also be represented in this way.
The basic primitives can be implemented in such a way that evaluation
at a point in time takes logarithmic time%
\footnote{Actually our implementation uses a bucketing heuristics that takes constant time on average.} 
and the other operations take
time linear in the number of line segments representing the inputs.

It seems that any exact time-dependent preprocessing technique needs a
basic ingredient that computes travel times not only for a point in
time a travel time \emph{profile}
but for an entire \emph{time-interval}.  An easy way to implement
this profile queury a generalization of Dijkstra's algorithm to profiles \cite{KS93}.
Tentative distances then become TTFs. Adding edge weights is replaced
by chaining TTFs and taking the minimum takes the minimum of
TTFs. Unfortunately, the algorithm looses its label-setting
property. However, the performance as a label-correcting algorithm
seems to be good in important practical cases.\frage{refs to refinements?}

\section{Construction}\label{s:construction}

The most expensive preprocessing phase of static CHs orders the nodes by importance.
For a first version we propose to adopt the \emph{static} algorithm for
the time-dependent CHs (TCHs).
This is based on the assumption that averaged over the planning
period, the importance of a node is not heavily affected by its
exact traffic pattern. 

The second stage of CH-preprocessing -- contraction -- is in principle
easy to adapt to time-dependence: we \emph{contract} the nodes of the
graph in the order computed previously. When contracting node $v\in
V'$, we are given a current (time-dependent) overlay graph
$G'=(V',E')$. For every combination of incoming edge $(u,v)\in E'$ and
outgoing edge $(v,w)\in E'$ we have to decide whether the path
$\seq{u,v,w}$ may be a shortest path at any point in time.  If so, we
have to insert the shortcut $(u,w)$ into the next overlay graph
$G''=(V'\setminus\set{v}, E'')$. The weight function of this shortcut can be computed
by chaining the weight functions of its constituents.
Later, we only need to consider shortcuts during time intervals when they
may represent a shortest path.\footnote{Although this can be viewed as a violation of
the FIFO-property, we do not get a problem when appliying time-dependent Dijkstra --
it never makes sense to wait for a shortcut to become valid since
this would not result in a shortest connection.}
The
required information can be computed by running profile-Dijkstra from
each node $u$ with $(u,v)\in E'$. The shortcut is needed for $w$ if
$c((u,v)\concat (v,w))<d(u,w)$ at any point in time.

\section{Query}\label{s:query}

The basic static query algorithm for CHs consists of a forward search in
an upward graph $\Gup=(V,\Eup)$ and a backward search in a downward graph $\Gdown$.
Wherever, these searches meet, we have a candidate for a shortest path.
The shortest such candidate is a shortest path.

Since the departure time is known, the forward search is easy to
generalize. In particular, the only overhead compared to the static case
is that we have to evaluate each relaxed edge for one point in time.
In our experience with a plain time-dependent Dijkstra, this means
a small constant factor overhead in practice.

The most easy way to adapt the backward search is to explore
\emph{all} nodes that can \emph{reach} $t$ in $\Gdown$.
Experiments for static CHs \cite{Gei08} indicate that
this search space is only a small constant factor larger than
the search space that takes edge weights into account.
During this exploration we mark all edges connecting nodes that
can reach $t$. Let $\Emarked$ denote the set of marked edges.

Now, we can perform an $s$--$t$-query by a forward search from $s$ in
$(V, \Eup\cup\Emarked)$. 

\begin{theorem}
The above algorithm is correct.
\end{theorem}
\begin{proof}(Outline)
This immediately follows from the properties of TCHs.
The detailed proof is analogous to the proof in \cite{Gei08}. 
Roughly, the properties of TCHs imply that there
must be a shortest path $P$ in the TCH that consists of two segments:
One using only eges in $\Gup$ leading to a peak node $v_p$ and
one connecting $v_p$ to $t$ in $\Gdown$. Since all edges of $P$ are
in the search space of our forward search, this path or some other
shortest path will be found.  
\end{proof}

\section{Refinements}\label{ss:crefinements}

\subsection{Node Ordering}\label{ss:orefinements}

Note that there are many ways to adapt the
node ordering to take time-dependence into account without resorting
to full-fledged time-dependent processing. For example, we can take
the average travel time of an edge or look at a sample of departure times
and base our priority for node-ordering on the entire sample.

\subsection{Contraction}\label{ss:crefinements}

The main difficulty in constructing TCHs is that 
the the complexities of time-dependent edge weights and
tentative distances grows with progressive contraction
and with the diameter of the profile-Dijkstra searches.
One way to counter this is to use approximations.
With some care, this can be done without compromising 
the exactness of queries.
In particular, we propose to compute piece-wise linear
approximations that are always within a factor $1+\epsilon$ from the
true travel time.

First, during a local search, we can replace tentative distances with less complex
upper bounds on the tentative distance.
The worst that can happen is that we introduce additional shortcuts.
The hope is that for sufficiently good approximations of the true
tentative distance, the number of superfluous shortcuts will be small.
The intuition behind this is that if traffic changes the shortest path at all,
it is unlikely that the travel time difference is tiny.  

For shortcuts that are actually introduced, we compute both upper and lower bounds.
For comparing a shortcut $a$ with a witness $b$, we compare a lower bound for $a$ 
with an upper bound for $b$. 
Once the (approximate) TCH is computed, we have a choice whether we 
want to condense it into an exact TCH (i.e., for all shortcuts
introduced, we compute there exact edge cost functions)
or we later modify the query to compute exact shortest paths using
approximate TCHs (ATCH). Note that
the complexity of the functions affects the space requirements but
has little influence on the cost of evaluation and thus on the query time.

\subsection{Query}\label{ss:qrefinements}

We can prune the forward search by marking all nodes $v$ in the 
backward search space with a lower bound $\ell(v)$ on the travel time to $t$.
Note that this information can be gathered with a static Dijkstra algorithm
that is likely to be faster than time-dependent Dijkstra.
Furthermore, we compute an upper bound $U$ for the travel time from $s$ to $t$
using any static routing technique, unpacking of the statically optimal path $P$,
and time-dependent evaluation of $P$.
Now, during forward search, if $d(s,v)+\ell(v) > U$ we do not need to continue
the search. 

There are various ways to compute better upper and lower bounds.
Assume we have computed a lower bound $L$ on the total travel time
using search in a static graph. Using $U$, $L$ and the departure time,
we know a time window $W$ for the arrival time.
For computing the lower bounds $\ell(v)$ we can then perform 
a variation of Dijkstras algorithm that computes minimum
travel times over a time interval. If the time interval is small,
this might be fast.

\paragraph*{Exact Routing in ATCHs (Outline)}
We modify our query algorithm to compute a graph that contains all edges
that \emph{might} be in the shortest path tree using upper and lower bounds
in a conservative way. Then, using the pruning techniques from above,
we remove all parts of this graph that cannot be part of a shortest path
from $s$ to $t$ at a given departure time. 
Then, we unpack all surviving edges. Hopefully, the resulting
graph will mostly consist of a small number of partially overlapping
paths from $s$ to $t$.
Finally,
we perform an exact forward search from $s$ in the unpacked graph.

\section{Conclusions}\label{s:conclusions}

We have developed algorithmic ideas for time dependent routing using CHs.
Now experiments have to show whether already the most basic approach
or some of its refinements yields a good exact query algorithm for road networks
or public transportation.
If problems show up, it is likely that the density of the graph or
the complexity of shortcuts gets out of hands in the later stages
of contraction. From the experience with static routing
\cite{BDSSSW08}, it is likely that such problems could be mitigated using
a combination with goal directed techniques, e.g., arc-flags. 
Again from \cite{BDSSSW08} it could be expected that at least this 
combination will outperform SHARC \cite{Del08}.

For commercial applications, approximate queries are not a big problem.
In this case, many simplifications suggest themselves where we 
can simply use approximations of time dependent functions
that are neither upper nor lower bounds and where we only introduce
shortcuts that bring significant improvements. 


\bibliographystyle{splncs}
\bibliography{hwy}
\end{document}